\documentclass[prl,twocolumn,superscriptaddress]{revtex4}

\usepackage{graphicx}
 
\begin{document}

\title{A Novel Large Moment Antiferromagnetic Order in K$_{0.8}$Fe$_{1.6}$Se$_2$ Superconductor}

\author{Wei Bao}
\email{wbao@ruc.edu.cn}
\affiliation{Department of Physics, Renmin University of China, Beijing 100872, China}
\author{Q. Huang}
\affiliation{NIST Center for Neutron Research, National Institute of Standards
and Technology, Gaithersburg, MD 20899, USA}
\author{G. F. Chen}
\affiliation{Department of Physics, Renmin University of China, Beijing 100872, China}\author{M. A. Green}
\affiliation{NIST Center for Neutron Research, National Institute of Standards
and Technology, Gaithersburg, MD 20899, USA}
\affiliation{Dept. of Materials Science and Engineering, University of Maryland, College Park, MD 20742, USA}
\author{D. M. Wang}
\author{J. B. He}
\author{X. Q. Wang}
\affiliation{Department of Physics, Renmin University of China, Beijing 100872, China}
\author{Y. Qiu}
\affiliation{NIST Center for Neutron Research, National Institute of Standards
and Technology, Gaithersburg, MD 20899, USA}
\affiliation{Dept. of Materials Science and Engineering, University of Maryland, College Park, MD 20742, USA}

\date{\today}

\begin{abstract}
The discovery of cuprate high $T_c$ superconductors has inspired searching for unconventional superconductors in magnetic materials. A successful recipe has been to suppress long-range order in a magnetic parent compound by doping or high pressure to drive the material towards a quantum critical point, which is replicated in recent discovery of iron-based high $T_C$ superconductors \cite{rev2010l}. The long-range magnetic order coexisting with superconductivity has either a small magnetic moment or low ordering temperature in all previously established examples. Here we report an exception to this rule in the recently discovered potassium iron selenide \cite{C122924}. The superconducting composition is identified as the iron vacancy ordered K$_{0.8}$Fe$_{1.6}$Se$_2$ with $T_c$ above 30 K. A novel large moment 3.31 $\mu_B$/Fe antiferromagnetic order which conforms to the tetragonal crystal symmetry has the unprecedentedly high an ordering temperature $T_N\approx 559$ K for a bulk superconductor. Staggeredly polarized electronic density of states thus is suspected, which would stimulate further investigation into superconductivity in a strong spin-exchange field under new circumstance.
\end{abstract}


\maketitle

Adding to the excitement generated by the discovery of iron-based high $T_c$ superconductors in the past two years \cite{Kamihara2008,A033603,A033790,A042053,rev2010l}, a nominal K$_{0.8}$Fe$_2$Se$_2$ material recently has been reported to have $T_c$ above 30 K \cite{C122924}. Replacing K by Rb \cite{C125525} or Cs \cite{C123637} yields
isostructural superconductor at similarly high a $T_c$. Materials of nominally fully occupied K site with chemical formula (Tl,$A$)Fe$_x$Se$_2$ ($A$=K, Rb or Cs) are also superconductors with $T_c$ above 30 K when iron deficient composition $x\sim 1.8$ \cite{C125236}. Superlattice peaks at ($\frac{1}{5},\frac{3}{5},0$) and ($\frac{1}{4},\frac{3}{4},0$) in the notation of the tetragonal ThCr2Si2 structure have been observed in a transmission electron microscopy study of a 33 K potassium iron selenide superconductor, and they are attributed to Fe vacancy ordering \cite{D012059}.
Only the ($\frac{1}{5},\frac{3}{5},0$) type of superlattice peaks are present in our single crystal x-ray diffraction study on similar superconducting samples, indicating bulk Fe vacancy order at this wave vector \cite{D014882}. Chemical composition of the ``optimized'' sample, which has been used to detect superconducting gap in ARPES study \cite{C125980}, is determined to be charge balanced with the valence 2+ for Fe ions in the x-ray structure refinement. The newest family of Fe superconductors, therefore, turns out not to be heavily electron doped as previous nominal formulas imply. 

The ARPES studies indicate that these new iron selenide superconductors are fundamentally different from previously discovered LaFeAsO (1111), BaFe$_2$As$_2$, LiFeAs (111) and FeSe (11) families of iron pnictide and chalcogenide superconductors, in either lacking the hole Fermi surfaces at the center of the Brillouin zone \cite{C125980}  or having electronic ones \cite{D014556,D014923}, to challenge the applicability of the prevailing $s^{\pm}$ symmetry proposed for previous Fe-based superconductors \cite{A032740,A033325}. The superconducting gap of 8-10 meV is isotropic in plane\cite{C125980,D014556,D014923} and NMR investigation indicates a singlet for the Cooper pair \cite{D011017}. These experiments put strong constraint on possible symmetry of the superconducting order parameter. More zone center phonon modes than the tetragonal ThCr2Si2 structure would allow have been observed in optical \cite{D010572} and Raman scattering \cite{D012168} measurements, corroborating an enlarged crystallographic unit cell \cite{D014882} to generate more optical phonon modes.  However, there are conflicting reports as to the existence of long-range magnetic order in the superconducting state from $\mu$SR and NMR works \cite{D011873,D014967}.

\begin{figure}
\includegraphics[width=60mm]{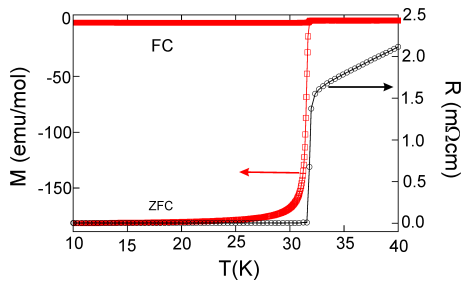}
\vskip -.2cm
\caption{Bulk superconducting transition of the K$_{0.8}$Fe$_{1.6}$Se$_2$ sample.
Magnetization measured with the sample cooled in zero magnetic field (ZFC) and in the 10 Oe measurement field (FC) is shown with resistivity.
}
\label{fig0}
\end{figure}

Here we report a neutron diffraction study on K$_{0.8}$Fe$_{1.6}$Se$_2$ ($T_c\approx 32$ K, see Fig.~\ref{fig0}). The antiferromagnetic order and Fe vacancy order are determined in a $\sqrt{5}$$\times$$\sqrt{5}$$\times$1 unit cell. The ordered magnetic moment of 3.31 $\mu_B$/Fe at 11 K is the largest among all Fe pnictide and chalcogenide superconductor materials. Magnetic transition also occurs at a record high value of $T_N\approx 559$ K below an Fe vacancy order-disorder transition at $T_S\approx 578$ K. The superconducting transition has an obvious effect on magnetic order parameter, indicating interaction between the two coexisting long-range orders. Our results put the newest structural family of Fe superconductors in a unique class where high $T_c$ superconductivity and strong magnetic order coexist. Relation between these two long range orders in unconventional superconductors, which are generally viewed as competing order parameters or the magnetic one of which needs to go soft towards a quantum critical point, has to be re-assessed in light of this new discovery.

\begin{figure}
\includegraphics[width=85mm]{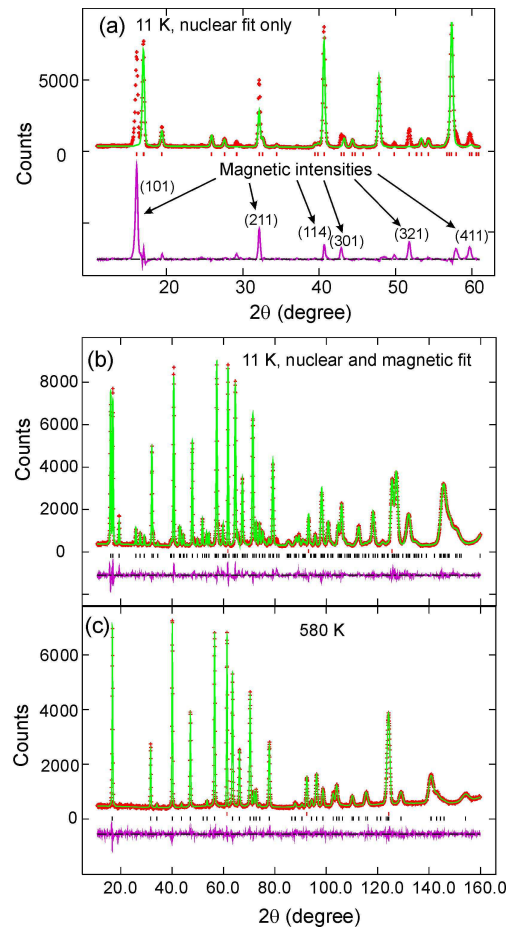}
\vskip -.2cm
\caption{Neutron powder diffraction spectra of the K$_{0.8}$Fe$_{1.6}$Se$_2$ superconductor. (a) Magnetic peaks are selected out of the 11 K spectrum. (b) The spectrum at 11 K is fitted by a Fe vacancy order described by space group $I4/m$ (No.\ 87) and an antiferromagnetic order by $I4/m^\prime$ in a tetragonal unit cell
of $a=b=8.6928(2)\AA$ and $c=14.0166(4)\AA$. Refer to Table I and Fig.~\protect\ref{fig2} for details.
(c) The spectrum at 580 K is described by the tetragonal ThCr$_2$Si$_2$ structure of lattice parameters $a=b=3.9289(1)\AA$ and $c=14.1906(8)\AA$. Both Fe vacancy and magnetic moment are disordered at the high temperature. Refinement parameters are listed in Table II. 
}
\label{fig1}
\end{figure}

\begin{figure*}
\includegraphics[width=170mm]{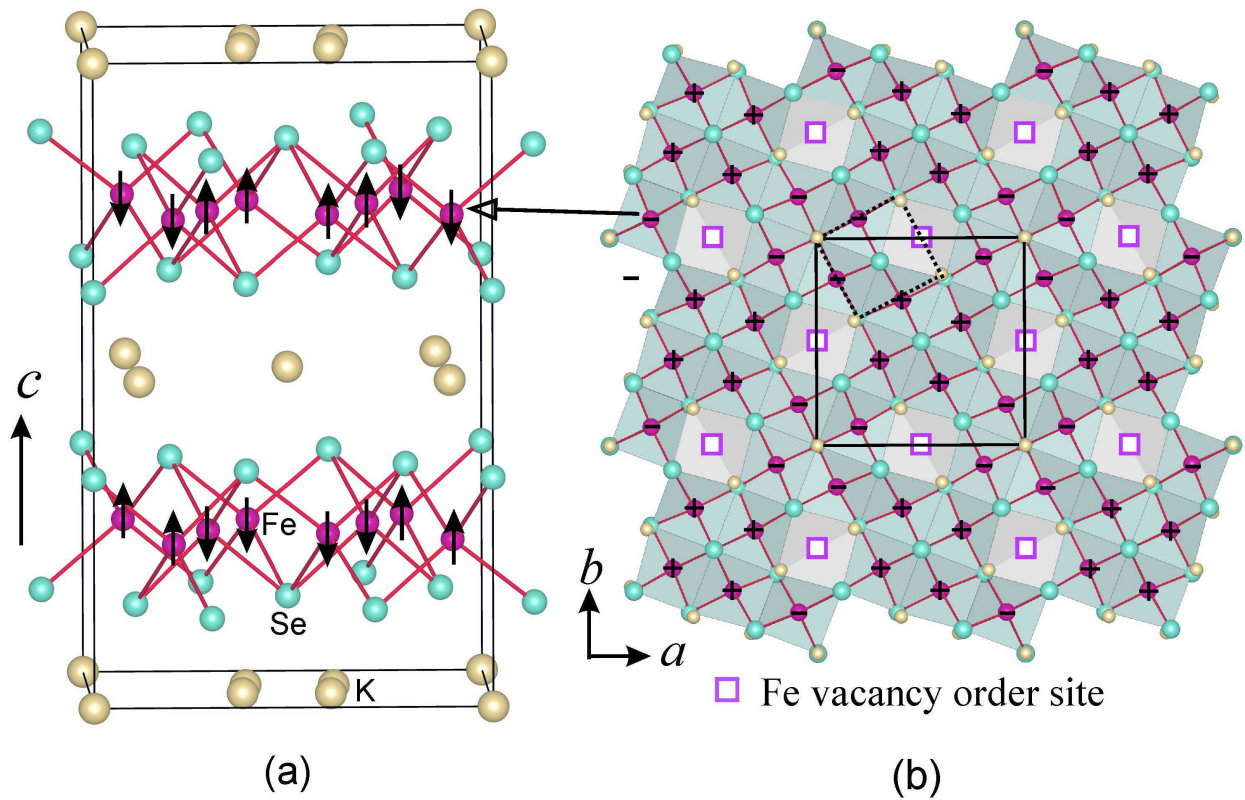}
\vskip -.2cm
\caption{Crystal and magnetic structure of K$_{0.8}$Fe$_{1.6}$Se$_2$ in the low-temperature $I4/m$ unit cell. (a) The top and the bottom Fe-Se layers, including magnetic moment orientation, form a mirror image of each other by the horizontal plane at $c/2$. (b) Top view of the top Fe-Se layer. The black solid line marks
the $I4/m$ unit cell.  
The iron vacancy site Fe(1) is marked by the open square, and the fully occupied Fe(2) site by purple circle with the $+$ or $-$ sign indicating magnetic moment direction which has only the c-axis component. The high-temperature $I4/mmm$ unit cell is marked by dash line, for which the low-temperature $I4/m$ unit cell is a $\sqrt{5}$$\times$$\sqrt{5}$$\times$1 supercell. There are two equivalent ways to generate from the high-temperature unit cell  the low-temperature supercell with different "handedness". Both of the handedness twins exist with equal probability in our sample according to our single crystal diffraction study.}
\label{fig2}
\end{figure*}

Neutron powder diffraction spectra were measured at various temperatures from 11 to 580 K.
Fig.~\ref{fig1}(b) shows the spectrum at 11 K with fitting curve from the Rietveld refinement. The small $2\theta$ portion of the spectrum is highlighted in Fig.~\ref{fig1}(a). The nuclear neutron diffraction peaks from the Fe vacancy ordered crystal structure, which have been identified in our previous x-ray diffraction work \cite{D014882}, are marked by green fitting curve in the spectrum at top. Remaining neutron diffraction peaks shown below are due to antiferromagnetic long-range order.
The refined crystalline and magnetic structure is presented in Fig.~\ref{fig2}, and the refinement parameters are listed in Table I. 
The sample composition determined from the refinement at 11 K `is K$_{0.82(2)}$Fe$_{1.626(3)}$Se$_2$. 
Thus, Fe ion in this sample also has 2+ valance. Either the heavy electron doping or valance fluctuation, which has served as staring point in some physics discussions on the new Fe selenide superconductors in current literature basing on incorrect nominal compositions, is misguided.

Each unit cell of K$_{0.8}$Fe$_{1.6}$Se$_2$ contains a pair of the Fe-Se layers related by inversion symmetry, see Fig.~\ref{fig2}(a). Top view of the Fe-Se layer is shown in Fig.~\ref{fig2}(b). The Fe vacancy occupies the $4d$ site marked by the open square. The Fe(2) site $16i$ (purple circle) is fully occupied. Thus, there is no Fe site disorder in stoichiometric K$_{0.8}$Fe$_{1.6}$Se$_2$ in the ordered structure,
in complete agreement with narrow NMR linewidth \cite{D011017,D014967}. 
Each Fe(2) ion carries a saturated magnetic moment 3.31(2) $\mu_B$ at 11 K, larger than the previous record of 2 $\mu_B$ per Fe found in parent compounds of Fe-based superconductors \cite{A092058,FeTe_comm}. The Fe magnetic moments form a collinear antiferromagnetic structure with the c-axis as the magnetic easy axis. The magnetic unit cell is the same as the crystalline one, and the magnetic ordering vector is {\bf Q}$_m$=(101). In an important difference from all previous 1111 \cite{A062195,A040795}, 122 \cite{A062776}, 11 \cite{A092058} and 111 \cite{B050525} families of Fe-based superconductor materials, antiferromagnetic order in K$_{0.8}$Fe$_{1.6}$Se$_2$ occurs in a tetragonal unit cell, maintaining the four-fold symmetry. This may have important ramification in current theoretical discussions concerning nematic order in high $T_c$ iron superconductors.

Being commensurate with crystal structure, the antiferromagnetic order in K$_{0.8}$Fe$_{1.6}$Se$_2$ is not as complex as the incommensurate one in Fe$_{1+x}$Te ($x>0.14$) \cite{A092058}, but it is considerably more complex than the two kinds of commensurate antiferromagnetic in-plane orders observed so far in iron pnictide and chalcogenide parent compounds, for which magnetic moments lie parallel along the shorter $b$-axis \cite{A062195,A062776,A092058,B050525}. An convenient way to comprehend the antiferromagnetic structure of K$_{0.8}$Fe$_{1.6}$Se$_2$ is to consider the four parallel magnetic moments at the center or the corners of the unit cell as a supermoment, Fig.~\ref{fig2}(b). The supermoments then form a simple chess-board nearest-neighbor antiferromagnetic order on a square lattice. It is interesting to note that along the edge of the primitive Fe square lattice, the sign of a string of four Fe(2) moments, terminated by the empty Fe(1) site, follows the same $++--$ pattern as found in the commensurate antiferromagnetic order in Fe$_{1+x}$Te ($x\le 0.076$) \cite{A092058,FeTe_comm}.
This observation suggests similar magnetic interaction in the $ab$-plane as in the 11 antiferromagnets. Along the c-axis, antiferromagnetic coupling prevails in K$_{0.8}$Fe$_{1.6}$Se$_2$ as well as  in LaFeAsO \cite{A040795}, 122 \cite{A062776}, 11 \cite{A092058} and 111 \cite{B050525} antiferromagnetic compounds, unless the rare earth magnetic ordering alters the coupling to become a parallel one \cite{A062195}. It would be interesting to investigate how our observed antiferromagnetic order is stabilized against those proposed in theoretical study \cite{C126015}.

\begin{figure}
\includegraphics[width=72mm]{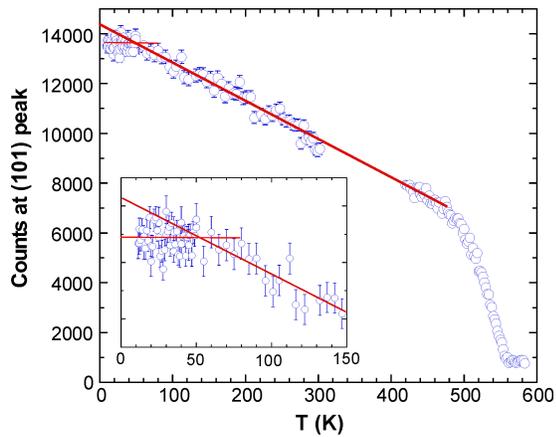}
\vskip -.2cm
\caption{The squared magnetic order parameter. The magnetic Bragg peak (101) appears below the N\'{e}el temperature $T_N\approx 559$ K. Below $\sim$500K, the Bragg intensity decreases linearly with temperature, then saturates below $\sim$2$T_c$. The inset enlarges the low temperature part around $T_c$.}
\label{fig3}
\end{figure}

The most surprising aspect of the large moment antiferromagnetic order in K$_{0.8}$Fe$_{1.6}$Se$_2$  is
its coexistence with the high $Tc\approx 32$ K superconductivity. 
Fig.~\ref{fig3} presents magnetic Bragg peak (101) as a function of temperature. The N\'{e}el temperature $T_N\approx 559$ K is astonishingly high for a superconductor, corroborating a recent report of a lower yet very high $T_N\approx 477$ K in a
closely related superconductor of nominal composition Cs$_{0.8}$Fe$_{2}$Se$_{1.96}$ from a $\mu$SR study \cite{D011873}. In a wide temperature range below $\sim$500 K, the squared magnetic order parameter decreases linearly with temperature (Fig.~\ref{fig3}). However, it departs from the trend and saturates at a constant value when $T_c$ is approached. The anticipation of the incipient superconducting transition indicates interaction between magnetism and superconductivity in the iron selenide superconductor.

\begin{figure}
\includegraphics[width=82mm]{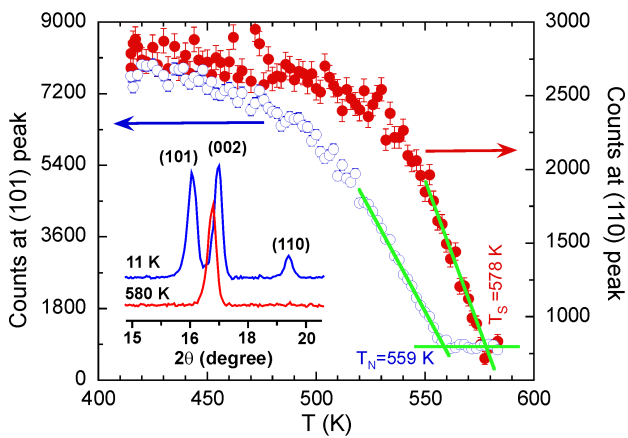}
\vskip -.2cm
\caption{Magnetic order depends on Fe occupancy order. The magnetic Bragg peak (101) and structural peak (110) as a function of temperature, showing magnetic transition at $T_N=559$ K and structural transition at $T_S=578$ K. Inset: Powder diffraction spectra at 11 and 580 K. The (002) exists in both the $I4/mmm$ and $I4/m$ phases. The appearance of (110) signifies the reduction of the $I4/mmm$ crystalline symmetry to the $I4/m$. The (101) develops with the antiferromagnetic order on the sublattice of Fe(2) ions.}
\label{fig4}
\end{figure}

The order-disorder transition of Fe vacancy occurs at $T_S\approx 578$ K, about 20 K above $T_N$, see Fig.~\ref{fig4}.
Unlike the magnetic (101) peak, structural Bragg peak (110) due to Fe vacancy order saturates towards its low temperature value rapidly. At room temperature, there is little difference in the (110) intensity, henceforth in the vacancy order, from that at 11 K, see Table I.  
The disappearance of the (110) peak above $T_S$ signifies the equivalence of the Fe(1) and Fe(2) sites and the restoration of the $I4/mmm$ symmetry. 
The neutron powder diffraction spectrum at 580 K shown in Fig.~\ref{fig1}(c) was fitted using the ThCr$_2$Si$_2$ structure with parameters listed in Table II. The iron site is  80.5(4)\% occupied. A lower $T_N$ than $T_S$ in K$_{0.8}$Fe$_{1.6}$Se$_2$ confirms our expectation that 
the establishment of the antiferromagnetic order depends on the development of the Fe occupancy order. However,
a perfect Fe order is not a prerequisite condition, as at 550 K which is only 9 degree below $T_N$, the 
Fe(2) occupancy is 93.6(3)\% (Table I). 

The large magnetic moment 3.31 $\mu_B$/Fe of K$_{0.8}$Fe$_{1.6}$Se$_2$ suggests strongly the possibility of a spin-polarized density-of-states at Fermi level. Superconductivity at $T_c\approx 32$ K develops in such a strong staggered polarizing electronic environment. Superconductivity in strong magnetic field was investigated theoretically by
Fulde and Ferrell \cite{fflo1} and Larkin and Ovchinnikov \cite{fflo2} (FFLO) four decades ago, and more recently superconductor at very high magnetic fields was reported \cite{uji,balicas}. A new mechanism for superconductivity which generates the FFLO scenario in a uniform magnetic field to a staggered internal magnetic field excerted by the large moment antiferromagnetic order is called for.
Together with different electron band structure revealed in ARPES studies \cite{C125980,D014556,D014923}, the new iron selenide superconductors may be regarded as a different class from previous Fe-based superconductors. From this physics perspective, even if signatures of unconventional superconductivity of the $s^{\pm}$ symmetry do not materialize, it may not lead to automatic disqualification of the theory for the former superconductors. 

In summary, the bulk superconductor K$_{0.8}$Fe$_{1.6}$Se$_2$ is charge balanced. The condensation of the Fe$^{2+}$ ions onto the Fe(2) site at $T_S\approx 578$ K leads to a well ordered stoichiometric crystal structure in $I4/m$ symmetry. A novel antiferromagnetic order with large saturated moment 3.31(2) $\mu_B$ per Fe develops on the vacancy decorated lattice below $T_N\approx 559$ K. Remarkably, superconductivity at a high $T_c\approx 32$ K realizes in such a strong magnetic environment. A new avenue to magnetic high $T_c$ superconductivity is open for both experimental and theoretical explorations.

\textbf{Methods}
K$_{0.8}$Fe$_{1.6}$Se$_2$ single crystals were synthesized from starting materials containing K, Fe and Se in the ration 0.8:2.3:2 using the Bridgeman method \cite{D010789}.
Resistivity was measured by the standard 4-probe method and magnetic susceptibility was measured in a 10 Oe magnetic field using a Physical Property Measurement System (PPMS) made by Quantum Design for several pieces from each batch of single crystals to monitor and ensure sample uniformity. Bulk superconductivity below $T_c\approx 32$ K is demonstrated by diamagnetic response in magnetic susceptibility as well as the $\lambda$ anomaly in specific heat measurement. 
10 g of crystals were ground to powders in a Helium atmosphere and then sealed with He exchange gas in a vanadium can. Sample temperature below room temperature was controlled by a He closed cycle refrigerator, above room temperature by a vacuum furnace.
Neutron powder diffraction experiments were carried out using the high resolution
powder diffractometer BT1 at the NIST Center for Neutron Research (NCNR). Cu(311) and Ge(311) monochromators were used to produce monochromatic neutron beams of wavelength 1.5403 $\AA$ and 2.0783 $\AA$, respectively. Collimators with horizontal divergences of 7', 15' (or 60') and 20' were used after the sample, before and after the monochromator, respectively.
The intensity was measured in steps of 0.05$^o$ in the 2$\theta$ range 3$^o$-168$^o$.
The nuclear and magnetic structures were refined using the GSAS program.

\textbf{Acknowledgments}:
We thank Q. M. Zhang, Z. Y. Lu and W. Q. Yu for useful discussions.
The work at RUC was supported by the National Basic Research Program of China (973 Program) under Grant No. 2011CBA00112, 2010CB923000 and 2009CB009100, and by the National Science Foundation of China under Grant No. 11034012, 10834013 and 10974254. 




\begin{thebibliography}{33}

\bibitem{rev2010l}
D.~C. Johnston,
  Adv. in Phys. \textbf{59},
  803 (2010).

\bibitem{C122924}
J.~Guo et al., Phys. Rev. B \textbf{82},
  180520(R) (2010).

\bibitem{Kamihara2008}
Y.~Kamihara et al., J.\ Am.\ Chem.\ Soc. \textbf{130},
  3296 (2008).

\bibitem{A033603}
X.~H. Chen et al.,
   Nature  \textbf{453},
  761 (2008).

\bibitem{A033790}
G.~F. Chen et al.,
   Phys. Rev. Lett.  \textbf{100},
   247002  (2008).

\bibitem{A042053}
Z.~A. Ren et al.,
   Chinese Phys. Lett.  \textbf{25},
  2215  (2008).

\bibitem {C125525}
 A.~F. Wang et al.,
   arXiv:1012.5525.

\bibitem {C123637}
 A. Krzton-Maziopa et al.,
   arXiv:1012.3637.

\bibitem{C125236}
M.~Fang et al.,
   arXiv:1012.5236.

\bibitem{D012059}
 Z. Wang et al., arXiv:1101.2059.

\bibitem{D014882}
P. Zavalij et al.,
   arXiv:1101.4882.
  

\bibitem{C125980}
Y. Zhang et al.,
  arXiv:1012.5980.
  

\bibitem{D014556}
D. Mou et al.,
  arXiv:1101.4556.

\bibitem{D014923}
 X.-P. Wang et al.,
   arXiv:1101.4923.  

\bibitem{A032740}
I.~I. Mazin et al., Phys. Rev. Lett. \textbf{101},
  057003 (2008).

\bibitem{A033325}
K. Kuroki et al.,
   Phys. Rev. Lett.  \textbf{101},
   087004  (2008).

\bibitem{D011017}
W. Yu et al.,
   arXiv:1101.1017.

\bibitem {D010572}
 Z.~G. Chen et al.,
   arXiv:1101.0572. 

\bibitem{D012168}
 A.~M. Zhang etal.,
   arXiv:1101.2168. 

\bibitem{D011873}
Z. Shermadini et al.,
   arXiv:1101.1873.

\bibitem{D014967}
Torchetti et~al., arXiv:1101.4967.

\bibitem{A092058}
Bao et~al., Phys. Rev. Lett.
  \textbf{102}, 247001
  (2009).

\bibitem{FeTe_comm}
Fruchart et~al., Mater. Res. Bull.
  \textbf{10}, 169 (1975).

\bibitem{A062195}
Qiu et~al., Phys. Rev. Lett. \textbf{101},
  257002 (2008).

\bibitem{A040795}
C. de~la Cruz et al.,
   Nature
  \textbf{453}, 899 (2008).

\bibitem{A062776}
Q. Huang et al.,
  Phys. Rev. Lett. \textbf{101},
   257003  ( 2008).

\bibitem {B050525}
 S. Li et al.,
  Phys. Rev. B  \textbf{ 80},
   020504(R) (2009).

\bibitem{C126015}
X.-W. Yan et al.,
   arXiv:1012.6015.

\bibitem {fflo1}
P. Fulde and R.~A. Ferrell,
  Phys. Rev. A \textbf{ 135},
  550  (1964).

\bibitem{fflo2}
A.~I. Larkin and Y.~N. Ovchinnikov ,
   Sov. Phys. JETP  \textbf{ 20},
   762  (1965).

\bibitem{uji}
S. Uji et al.,
  Nature  \textbf{ 410},
   908  ( 2001).

\bibitem{balicas}
 L. Balicas et al.,
   Phys. Rev. Lett.  \textbf{ 87},
  067002  (2001).

\bibitem{D010789}
 D.~M. Wang et al.,
   arXiv:1101.0789.

\end{thebibliography}
\end{document}